\documentclass[aps,prl,multicol]{revtex}

\def\gsim{ \lower .75ex \hbox{$\sim$} \llap{\raise .27ex \hbox{$>$}} }
\def\lsim{ \lower .75ex \hbox{$\sim$} \llap{\raise .27ex \hbox{$<$}} }

\begin{document}

\title{Conditions for Generating Scale-Invariant Density Perturbations}

\author{Steven Gratton$^1$, Justin Khoury$^{1,2}$, Paul J. Steinhardt$^1$ and Neil 
Turok$^3$}

\address{$^1$Joseph Henry Laboratories,
Princeton University,
Princeton, NJ 08544, USA \\
$^2$Institute for Strings, Cosmology, and
Astroparticle Physics, Columbia University, New York, NY 10027, USA \\
$^3$ DAMTP, CMS, Wilberforce Road, Cambridge, CB3 0WA, UK}

\maketitle

\begin{abstract}
We  analyze
the general conditions on the equation of state
$w$ required for  quantum fluctuations of a
scalar field to produce  a scale-invariant spectrum of density
perturbations, including models which (in the four dimensional
effective description) bounce from a contracting to an expanding
phase.
We show that there are   only two robust cases:  $w\approx -1$
(inflation) and $w \gg 1$ (the ekpyrotic/cyclic
scenario).  All other cases, including the $w \approx 0$ case
considered by some authors, require extreme fine-tuning of
initial conditions and/or the effective potential.
For the ekpyrotic/cyclic ($w \gg 1$) case, 
we also analyze the  small deviations from
scale invariance.
\end{abstract}

\begin{multicols}{2}[]

Until recently, the only known mechanism for generating super-horizon,
adiabatic, scale-invariant density perturbations was
inflation~\cite{inf}.  One of the great surprises of the
ekpyrotic~\cite{orig,seiberg,pert} and cyclic models~\cite{st,khoury2}
is that they provide a second, alternative mechanism.  In this paper,
we introduce a gauge invariant, systematic analysis (including
gravitational backreaction) that identifies the most general conditions
required assuming the density perturbations arise as a result of
quantum fluctuations of a single scalar field $\phi$ with potential
$V(\phi)$.  We include the possibility that the
universe may bounce from a contracting to an expanding phase, and that
the perturbations can be matched across such a
bounce in an unambiguous way~\cite{seiberg,pert,tolley,tts}.

First, we must obtain scale-invariant fluctuations during a period in
which the scalar field dominates the energy density of the universe.
The conditions for this to occur can be characterized by the
equation of state $w$ during this epoch.  We find 
three interesting cases,
for each of which  
 $w$ is  nearly constant: $i)$ an expanding universe with $w
\approx -1$, corresponding to slow-roll inflation; $ii)$ a {\it
contracting} universe with $w\gg 1$, corresponding to the
ekpyrotic/cyclic models; and, $iii)$ a {\it contracting} universe with
$w\approx 0$, as discussed by Wands~\cite{wands}  and by Finelli and
Brandenberger~\cite{bran}.  Although the last case does generate
a scale invariant spectrum of curvature perturbations,
we shall show that the corresponding 
 Newtonian potential has a very red power spectrum. 
This points to
a serious instability of the background, which 
we explicitly identify in the infinite wavelength limit.
Unlike the ekpyrotic/cyclic cases, the $w=0$ 
background solution is unstable and hence not a dynamical
attractor.  The only additional cases
are ones in  which $w$ is rapidly time-varying, but these
require extreme fine-tuning of $V(\phi)$.

Hence, we  find that
 the ekpyrotic/cyclic case
remains as the only viable alternative
to inflation, assuming successful matching of the 
growing mode perturbations generated during the
contracting phase onto growing mode 
perturbations in the expanding phase, as proposed in
Refs.~\ref{pert}, \ref{tolley}, and \ref{tts}.
Using this prescription, we 
derive an expression 
for the deviation from scale-invariance for cyclic/ekpyrotic models
in terms of
``fast-roll'' parameters $\epsilon$ and $\eta$, analogous to the
``slow-roll'' parameters of inflation.

Our analysis is restricted to the case of a single scalar field, 
which includes the  simplest inflationary scenarios as well as the
ekpyrotic/cyclic models.  
In Newtonian gauge, the perturbed metric for a spatially-flat
background can be expressed in terms of a single gauge invariant
variable $\Phi$, the Newtonian potential, as
\begin{eqnarray}
\nonumber
& & ds^2 = a^2(\tau)\cdot\{-(1+2\Phi(\vec{x},\tau))d\tau^2 + (1-
2\Phi(\vec{x},\tau))d\vec{x}^2\}\,,
\end{eqnarray}
where $\tau$ is conformal time, and where we have used the fact that
fluctuations of a scalar field do not generate anisotropic stress.

While knowledge of $\Phi$ is sufficient to determine the perturbed
metric, it is useful to introduce a second variable,
$\zeta$, which is the curvature perturbation on comoving
hypersurfaces~\cite{bardeen,BST}.
$\Phi$ and
$\zeta$ are related by
\begin{equation}
\zeta = \frac{2}{3a^2(1+w)}\left(\frac{\Phi}{a'/a^3}\right)'\,,
\label{eq:zeta}
\end{equation}
where a prime denotes differentiation with respect to $\tau$.
The variable $\zeta$
has the virtue of remaining nearly constant at  superhorizon wavelengths
during epochs of {\it expansion}.  
In inflationary models, it allows one to easily match the Newtonian potential
at horizon reentry in the matter dominated phase to that calculated at
horizon exit during the inflationary phase.  In the ekpyrotic/cyclic
scenarios, once its spectrum has been determined {\it after the bounce
to an expanding phase}, 
$\zeta$ also 
gives the perturbation amplitude at horizon re-entry.

It is tempting to suppose that we should only be
interested in tracking the evolution of $\zeta$.  However, even 
though $\zeta$ is continuous throughout a contracting phase or an expanding
phase, it can undergo a rapid jump during the transition between
the two \cite{pert,tts}.
It is necessary to match the incoming 
$\Phi_{in}$ and $\zeta_{in}$ to the outgoing 
$\Phi_{out}$ and $\zeta_{out}$.  Generically there is some mixing 
and $\zeta_{out}$ 
depends on  a  combination of the 
$\Phi_{in}$ and $\zeta_{in}$\cite{pert}. Hence, it is important
to know both $\Phi_{in}$ and $\zeta_{in}$ at the bounce.

For the ekpyrotic/cyclic case, 
$\Phi_{in} $ is scale-invariant and  $\zeta_{in}$ is blue (decreasing
at long wavelengths), and, hence, 
$\zeta_{out} $ 
is dominated by the scale-invariant
contribution (due to $\Phi_{in}$)
at long wavelengths,
leading to a scale-invariant spectrum
as modes re-enter the horizon during the expanding phase. 
For the $w\approx 0$ case, 
$\zeta_{in}$ is scale-invariant before the bounce and
$\Phi_{in}$ 
turns out to be red (increasing  
at long
wavelengths and exponentially larger than $\zeta_{in}$). 
This suggests that the background possesses
a serious long wavelength instability;  we confirm this 
by showing that the background  solution
is not a dynamical attractor.

{\it The $\Phi$ spectrum}: The differential equation for the $k$-mode
$u_k$ of the gauge invariant variable $u$, related to the Newtonian
potential by $u = a\Phi/\phi'$
(where we drop the $(in)$
subscript, henceforth), is
\begin{equation}
u_k'' + \left(k^2 - \frac{\beta\left(\tau\right)}{\tau^2} \right) u_k = 0\,;
\label{eq:u2}
\end{equation}
\begin{eqnarray}
\nonumber
 \beta\left(\tau\right) \equiv \tau^2 H^2 a^2 \left\{\bar{\epsilon} +
\frac{(1+\bar{\epsilon})}{2}\left(\frac{d\ln
\bar{\epsilon}}{dN}\right)\right. \\  
 \qquad\;\;\;\;\;\;\;\;\;\;\;\;\; \left. + \frac{1}{4}\left(\frac{d\ln
\bar{\epsilon}}{dN}\right)^2 - \frac{1}{2}\frac{d^2\ln
\bar{\epsilon}}{dN^2}\right\}\,, 
\label{m}
\end{eqnarray}
where $H=a'/a^2$ is the Hubble parameter, $N\equiv\ln a$, and
$\bar{\epsilon}\equiv 3(1+w)/2$.  Note that in the case of inflation,
$\bar{\epsilon}$ reduces to the usual ``slow- roll'' parameter, while
$N$ is the number of e-folds of expansion.

In the regime $k^2\tau^2\gg |\beta|$, Eq.~(\ref{eq:u2}) reduces to the
equation for a simple harmonic oscillator, and $u_k$ is stable.  When
$k^2\tau^2\ll |\beta|$, however, the amplitude of the mode is
unstable.  In order to have a situation where successive modes with
increasing $k$ are becoming unstable and growing, we need $\beta$
positive, and assuming that $\beta$ is slowly varying, we require that
$\tau$ be negative and increasing.  This applies to expanding models,
such as inflation, or contracting models, such as the ekpyrotic/cyclic
scenarios.

For general time-varying $w$, $\beta(\tau)$ will be a complicated function of
time, and one can use numerical methods to solve Eq.~(\ref{eq:u2}).
In the most plausible cases, however, it is reasonable to approximate
$w$ as constant, at least for the observationally relevant range of
modes.  

It is well-known~\cite{pert,lyth} that solutions with constant $w$
correspond to potentials of the exponential form, $V(\phi) =
-V_0e^{-c\phi}$, where $c$ and $V_0$ are constants.  In this case, the
equation of state is related to the slope of the potential by
\begin{equation}
\bar{\epsilon} \equiv \frac{3}{2}(1+w) = 
\frac{1}{2}\left(\frac{V_{,\phi}}{V}\right)^2\,,
\label{const}
\end{equation}
(we use units where $8\pi G=1$)
and the solution for the background, assuming homogeneity, isotropy
and spatial flatness, is given by
\begin{eqnarray}
a(\tau) \sim (-\tau)^{1/(\bar{\epsilon}-1)}&,& ~~
H = \frac{1}{(\bar{\epsilon}-1)a\;\tau} \nonumber \\ 
\phi' = \sqrt{\frac{2}{\bar{\epsilon}}} \left(\frac{\bar{\epsilon}} 
{\bar{\epsilon}-1}\right)\; \tau^{-1}      &,&~~
V(\phi) = -\frac{(\bar{\epsilon}-3)}{(\bar{\epsilon}-1)^2a^2\tau^2}.
\label{eq:exact}
\end{eqnarray}

Substituting the above into Eq.~(\ref{m}), we obtain
$\beta = \bar{\epsilon}/ (\bar{\epsilon}-1)^2\,$.
Since $\beta$ is constant in this case, Eq.~(\ref{eq:u2}) can be
solved analytically, with general solution
\begin{equation}
u_k=\sqrt{-k\tau}\, [C_1(k)J_n(-k\tau) + C_2(k)J_{-n}(-k\tau)]\,,
\label{eq:bessel}
\end{equation}
where $n\equiv \sqrt{\beta+1/4}$, $J_n$ is the Bessel function of the
first kind of order $n$, and $C_i(k), i=1,2$ are arbitrary functions
of $k$.

The functions $C_i(k)$ are determined by specifying initial conditions
when the mode is stable, {\it i.e.}, when $k^2\tau^2 \gg \beta$.  In
this limit, we make the usual assumption that the fluctuations in
$\phi$ are in their Minkowski vacuum, which corresponds to $u_k\approx
ie^{-ik\tau}/(2k)^{3/2}$.  Using the relation $u_k = a\Phi_k/\phi'$
and the asymptotic properties of Bessel functions, this gives
\begin{equation}
u_k=\frac{\sqrt{-\pi k\tau}}{4k^{3/2}\sin(\pi n)}
[J_{-n}(-k\tau)-e^{-i\pi n}J_{n}(-k\tau)]\,,
\label{eq:bessel2}
\end{equation}
where we have neglected an irrelevant phase factor. 

We are interested in the amplitude of the mode in the long-wavelength
regime, $k^2\tau^2\ll \beta$.  In this limit, we can expand the Bessel
functions to obtain
\begin{eqnarray}
\nonumber
& & k^{3/2}\Phi_k \approx\frac{\sqrt{\pi}}{2^{3/2}\sin(\pi n)\Gamma(1-
n)}\left(\frac{\phi'}{a}\right)\left(\frac{-k\tau}{2}\right)^{-n+1/2}\cdot\\
& & \left\{1-e^{-i\pi n}\frac{\Gamma(1-n)}{\Gamma(1+n)}\left(\frac{-
k\tau}{2}\right)^{2n} - \frac{\Gamma(1-n)}{\Gamma(2-n)}\left(\frac{-
k\tau}{2}\right)^{2}\right\}\,.
\label{eq:Phi}
\end{eqnarray}
The Newtonian potential has a scale-invariant spectrum if
the rms amplitude of $\Phi_k$ varies with $k$ as $k^{-3/2}$.  Hence,
we conclude that this will be the case if $n\approx 1/2$.  Recalling
that $n = \sqrt{\beta + 1/4}$,
 this can be
expressed as a constraint on $\beta$ or, equivalently,
$\bar{\epsilon}$:
\begin{equation}
\beta = \frac{\bar{\epsilon}}{(\bar{\epsilon}-1)^2}\ll 1\,.
\label{eq:condw}
\end{equation}

We have thus translated the requirement of scale invariance
for $\Phi$ into a condition on the background equation of state.
Therefore, we may now determine 
what choice of $w$
will satisfy Eq.~(\ref{eq:condw}) and  lead to a
Harrison-Zel'dovich spectrum.

First, this condition is clearly satisfied when $\bar{\epsilon} \ll
1$, that is, when $w \approx -1$.  This corresponds to the case of
slow-roll inflation~\cite{inf}.  Note, however, that there is a second
regime in which condition~(\ref{eq:condw}) holds, namely when
$\bar{\epsilon} \gg 1$, corresponding to $w \gg 1$.  This is the limit
relevant to the production of fluctuations in the ekpyrotic and cyclic
scenarios~\cite{orig,seiberg,pert,st,khoury2}.  These two regimes are
in some sense at opposite ends of parameter space.  In the
inflationary case, $\bar{\epsilon}$ plays the role of a slow-roll
parameter and is therefore small.  In the ekpyrotic and cyclic
scenarios, however, $\bar{\epsilon}$ is large compared to unity.
Also, from Eq.~(\ref{eq:exact}), we see that the universe is
expanding in the first case and contracting in the second.

This analysis  
 assumed  a nearly constant $w$
 so that 
$\beta(\tau)$ in Eq.~(\ref{m}) and, consequently, the spectral index
is nearly constant.
Note that this assumption is not necessary. 
It  is  possible, in principle,
 to build models for which the time-variation of $w$
is non-negligible,  and yet the derivative terms in Eq.~(\ref{m})
conspire to
cancel for a significant range of e-folds,  $N$.
This has
been discussed for inflation by Wang {\it et
al.}~\cite{wang}  who showed that 
maintaining the cancellation for many e-folds
requires
extreme fine-tuning of $V(\phi)$ compared to the constant
$w$ cases. 
Hence, these models seem highly unlikely.

{\it The $\zeta$ spectrum}: To calculate the spectrum of the second
gauge invariant variable of interest, $\zeta$, we substitute the
expression for $\Phi$ obtained in Eq.~(\ref{eq:Phi}) into
Eq.~(\ref{eq:zeta}).  The leading term in the expansion
for $\Phi$ is
\begin{equation}
k^{3/2}\Phi_k  \sim \left(\frac{\phi'}{a}\right)(-k\tau)^{-n+1/2}\,,
\end{equation}
where we have omitted the numerical coefficient.  As before, we 
approximate $w$  as constant, and, thus,
so is $n$.  Using Eqs.~(\ref{eq:exact}), we find
\begin{eqnarray}
\nonumber
& & k^{3/2}\Phi_k \sim k^{-n+1/2}(-\tau)^\lambda\;; \\
& & \lambda\equiv -\frac{1+\bar{\epsilon}}{2(\bar{\epsilon}-
1)}\left\{1+\frac{\bar{\epsilon}-1}{|\bar{\epsilon}-1|}\right\}\,,
\end{eqnarray}
as well as
\begin{equation}
\frac{a'}{a^3} \sim (-\tau)^{-(1+\bar{\epsilon})/(\bar{\epsilon}-1)}\,.
\end{equation}

Since the expression for $\lambda$ involves a factor of
$|\bar{\epsilon}-1|$, we must consider the cases $\bar{\epsilon}>1$
and $\bar{\epsilon}<1$ separately.  Starting with the latter (which
includes slow-roll inflation), we find $\lambda=0$, and therefore
\begin{equation}
k^{3/2}\left(\frac{\Phi_k}{a'/a^3}\right) \sim k^{-\bar{\epsilon}/(1-
\bar{\epsilon})} (-\tau)^{(1+\bar{\epsilon})/(\bar{\epsilon}-1)}\,.
\end{equation}
Using the relation between $\zeta$ and $\Phi$ given in
Eq.~(\ref{eq:zeta}), we see that $\zeta$ will have a scale-invariant
spectrum if $\bar{\epsilon}\ll 1$.  Recall that this limit corresponds
to slow-roll inflation and that it also leads to a scale-invariant
spectrum for $\Phi$.

For the case $\bar{\epsilon}>1$, which includes the ekpyrotic/cyclic
scenarios, we find $\lambda=-(1+\bar{\epsilon})/(\bar{\epsilon}-1)$,
and thus
\begin{equation}
k^{3/2}\left(\frac{\Phi_k}{a'/a^3}\right) \sim k^{-1/(\bar{\epsilon}-1)}\,.
\end{equation}
Since the right hand side is independent of time, Eq.~(\ref{eq:zeta})
implies that this leading term for $\Phi$ does not contribute to
$\zeta$.  In order to determine the long-wavelength piece of $\zeta$,
we must therefore keep the higher-order terms in the expansion for
$\Phi$ given in Eq.~(\ref{eq:Phi}).  It is straightforward to show
that the result is of the form
\begin{equation}
k^{3/2}\zeta \sim f_1(\tau)k^{\bar{\epsilon}/(\bar{\epsilon}-1)} + 
f_2(\tau)k^{(2\bar{\epsilon}-3)/(\bar{\epsilon}-1)}\,,
\label{expzeta}
\end{equation}
where $f_1(\tau)$ and $f_2(\tau)$ are time-dependent factors.

For the ekpyrotic/cyclic scenarios, corresponding to the regime
$\bar{\epsilon} \gg 1$, the first-term in Eq.~(\ref{expzeta}) gives
the dominant contribution at long-wavelengths, and thus $k^{3/2}\zeta$
goes like $k$.  Hence, while the condition $\bar{\epsilon}\gg 1$ led
to a scale-invariant spectrum for $\Phi$ in the pre-big bang phase, it
yields a blue spectrum for $\zeta$.  As we have seen above, this is a
consequence of the fact that the growing mode of $\Phi$, which is
scale-invariant in this limit, is projected out of $\zeta$.  Thus,
$\zeta$ is determined by the next-order correction in the expansion
for $\Phi$, which is down by a factor of $k^{2n}\approx k$.
Nevertheless, as mentioned earlier, $\zeta$ 
jumps at the bounce (mixes with the scale-invariant $\Phi$ mode)
and is dominated by the scale-invariant contribution at long 
wavelengths after the universe begins to expand.

Note that, if we choose $\bar{\epsilon}\approx 3/2$, then the second
term in Eq.~(\ref{expzeta}) dominates at large wavelengths, and the
resulting spectrum for $\zeta$ is nearly scale-invariant before
the bounce.  This case,
identified by Wands~\cite{wands} and recently studied by Finelli and
Brandenberger~\cite{bran}, describes a contracting universe with a
dust-like equation of state, $w\approx 0$.  However, using our
results 
above, we see that
the corresponding spectrum of $\Phi$ is strongly red ($k^{3/2} 
\Phi_k\sim k^{-2} \tau^{-5}$), indicating a severe long wavelength
instability.  
Then 
if $\Phi$ and $\zeta$ mix at all at the bounce, 
the red contribution would dominate at long wavelengths rendering
the resulting universe phenomenologically unacceptable. 

{\it Stability of the background solution:}
Before considering  evolution of quantum fluctuations,
we should first  consider whether the 
constant $w$ solutions we have assumed as background 
solutions are stable
attractors to the equations of motion.
If they are not, then they could not arise in a cosmological
solution without extraordinary fine-tuning of initial conditions.
The expanding inflationary ($w=-1$) is known to be a stable
attractor. Here we show that the contracting ekpyrotic/cyclic
($w\gg1$) phase is also a stable attractor, but the 
contracting $w=0$ phase is not.

The stability of the background solution may be studied
in the infinite wavelength ({\it i.e.,} homogeneous) limit simply by 
considering the scalar field equation in a homogeneous Universe:
\begin{equation}
\ddot{\phi} +3{\dot{a}\over a} \dot{\phi} = -V_{,\phi},
\label{phieq}
\end{equation}
with dots denoting derivative with respect to proper time $t$.
The Friedmann
equation allows us to express $\dot{a}/a$ in terms
of the scalar field. (For simplicity we do not perturb the
space curvature since the above calculation indicates
this effect is subdominant.) 
Setting $\phi=\phi_B +\delta \phi$, 
with $\phi_B$ the background ``scaling solution,''
\begin{eqnarray}
a(t) \propto (-t)^p, \quad  V(\phi_B) = -V_0 e^{-c \phi_B(t)} =p(3p-1)/t^2, 
\label{sc}
\end{eqnarray}
where $p=2/c^2$,
(\ref{phieq}) becomes
\begin{equation}
\delta \ddot{\phi} +{1+3p\over t} \delta \dot{\phi} - {1-3p\over t^2} \delta \phi =0,
\label{perts}
\end{equation}
whose two linearly independent solutions for $p \neq {2\over3}$
are
$\delta \phi \sim t^{-1}$ and $t^{1-3p}$. 
For all $p$, the first solution is just an infinitesimal
shift in the time to the big crunch: $\delta \phi \propto \dot{\phi}_B$.
Such a shift provides a solution to the
Einstein-scalar equations because they are time translation
invariant, but it is
physically irrelevant since it can be removed by
a redefinition of time. In contrast, 
however, the second solution represents
a physical perturbation of the background solution.
For $p>{1\over 3}$, the second solution grows
as $t$ approaches zero, indicating an instability of the
background scaling solution.  
(For $p=\frac{2}{3}$, the second solution is $t^{-1} \, {\rm ln}(-t)$.)
This is confirmed by
calculating the ratio of kinetic energy to potential
energy in the scalar field, which is constant and equal to
\begin{equation}
{K\over V}= \frac{1+w}{1-w}= \frac{1}{3p-1}
\label{kov}
\end{equation}
in the background solution. This ratio is 
unaltered by the first perturbation solution (since it is a time shift)
 but the second yields
\begin{equation}
\delta \left(K\over V\right) \propto (-t)^{1-3p}
\label{kovp}
\end{equation}
which for $p>{1\over 3}$ diverges as $t=0$ is approached.
This shows in particular that the $w=0$ 
($p={2\over 3}$) background scaling solution is unstable and hence
not an attractor.  Conversely,
the ekpyrotic/cyclic cases, which correspond
to $p<<1$, possess scaling solutions which are
are stable attractors in the infinite wavelength limit,
since the only growing mode is as we have discussed just
a time translation. 

A more general, if more heuristic argument can be obtained 
by comparing the Friedmann equation for the three cases,
\begin{equation}
H^2 \equiv \left(\frac{\dot{a}}{a}\right)^2 = \frac{1}{3} \rho
- \frac{k}{a^2},
\end{equation}
where $H$ is the Hubble parameter and $\rho$ is the energy density.
For the expanding $w=-1$ case, the
scalar field energy density is nearly constant, whereas the curvature,
radiation ($\propto 1/a^4$), matter ($\propto 1/a^3$), 
 and  other
forms of energy density 
all decrease   as $a$ expands. Hence, 
the contant energy density $w=-1$ state is an attractor.  
For the {\it contracting} $w \gg 1$ case
($a \sim t^p$ with $p \ll 1$), the scalar field energy density 
 increases as $1/a^{2/p}$ as $a$ decreases, whereas the curvature, 
matter, radiation, 
or other forms of energy density 
increase at a slower rate.
Hence, the $w\gg1$ contracting 
solution is also an attractor of the Friedmann
equation.   However, for the $w \approx 0$ solution, the 
scalar field energy density increases as $1/a^3$, but radiation density
 increases at a more rapid rate.  
Furthermore, as shown above, a small perturbation drives the universe
away from $w \approx 0$ ($K \approx V$) towards $w=1$ ($K \gg V$),
a state in which the energy density ($\propto 1/a^6$) increases
even more rapidly.
Hence, the $w \approx 0$ background  is
not a stable attractor.

We can state the conclusion more generally. 
For an exponential potential 
proportional to ${\rm exp}(-c \phi)$, we have
a scaling solution when 
$V$ is positive (negative)
for $c<\sqrt{6}$  ($c>\sqrt{6}$).
In the scaling solution, we have $w=2/(3p)-1= (c^2/3) -1$.
It follows from the analysis  above
that {\it in a contracting universe, the scaling
solution is only a stable attractor if $w>1$ (or $c>\sqrt{6}$)
and the scalar potential is negative}.
By 
time reversal, we infer that
{\it in an expanding Universe the scaling
solution is only a stable attractor if $w<1$ ($c<\sqrt{6}$) and 
the scalar potential is positive}. Inflationary and 
`quintessence'-type scaling solutions
are both included in this latter case.

Thus, we have completed our classification of the possibilities. In
particular, we have seen that when the universe is contracting with $w
\gg 1$, as in the ekpyrotic and cyclic scenarios, the Newtonian
potential $\Phi$ develops a scale-invariant spectrum while that of
$\zeta$ is blue.  
Provided $\Phi$ and $\zeta$ mix at the bounce
(which is to say that the growing mode of the contracting
phase does not match to a pure decaying mode in the expanding 
phase),  one obtains
scale-invariant density perturbations in the expanding phase. 
If any radiation is generated at the bounce, mixing is expected
\cite{pert,tts}.
When the universe is contracting with a dust-like
equation of state ($w\approx 0$), 
$\zeta$ acquires a scale-invariant spectrum, while $\Phi$ 
acquires a red spectrum. With mixing  at the
bounce, one  obtains an unacceptable red spectrum.
More generally, we have shown that the 
$w=0$ scaling solution is not an attractor.
Therefore,
there is no reason to expect the universe to reach the scaling solution
in the first place.
 More generally, we have shown that
the only way to obtain a stable attractor scaling background
solution in
a contracting universe is to have 
a {\it negative} scalar field potential, as in the
ekpyrotic and cyclic models.

{\it Spectral index in ekpyrotic and cyclic models}: In the remaining
part of this paper, we shall focus on the ekpyrotic/cyclic generation
of perturbations and calculate the spectral index, giving
a treatment analogous to that given for inflation
by Wang {\it et al.}~\cite{wang}.

Recall that approximate scale invariance of the power spectrum in the
ekpyrotic/cyclic scenario requires that $\bar{\epsilon}$ be large and
nearly constant as modes become unstable.  Since $\bar{\epsilon}\gg
1$, it is convenient to introduce a small, ``fast-roll'' parameter
$\epsilon$ as
\begin{equation}
\epsilon\equiv \frac{1}{2\bar{\epsilon}} =
\left(\frac{V}{V_{,\phi}}\right)^2\,. 
\label{neweps}
\end{equation}
The condition $\bar{\epsilon}\gg 1$ implies $\epsilon\ll 1$, which
translates into the requirement that the potential be steep.

Since $\bar{\epsilon}$ is large and nearly constant, the parameter
$\beta$ defined in Eq.~(\ref{m}) reduces to
\begin{equation}
\beta \approx \tau^2H^2a^2\bar{\epsilon}\left\{1 + 
\frac{1}{2}\left(\frac{d\ln\bar{\epsilon}}{dN}\right) \right\}\,,
\label{m2}
\end{equation}
where we have assumed that $d^2\ln\bar{\epsilon}/dN^2$ and $
(d\ln\bar{\epsilon}/dN)^2$ are much smaller than $d\ln\bar{\epsilon}/dN$.

Recalling from Eq.~(\ref{const}) that $\bar{\epsilon}\approx
V_{,\phi}^{\,2}/2 V^2$ for nearly constant $\bar{\epsilon}$, we obtain
\begin{equation}
\frac{d\ln\bar{\epsilon}}{dN} = 
\left(\frac{\phi'}{aH}\right)\frac{d\ln\bar{\epsilon}}{d\phi} = -
2\left(\frac{V_{,\phi}}{V}\right)\left(\frac{\phi'}{aH}\right)\eta\,,
\label{eq:dlneps}
\end{equation}
where we have introduced a second fast-roll parameter $\eta$, defined by
\begin{equation}
\eta \equiv  1 - \frac{VV_{,\phi\phi}}{V_{,\phi}^2}\,.
\end{equation}
Note that $\eta=0$ corresponds to pure exponential potentials.

Substituting for $\phi'/(aH)$ using Eqs.~(\ref{eq:exact}),
Eq.~(\ref{eq:dlneps}) reduces to
\begin{equation}
\frac{d\ln\bar{\epsilon}}{dN} \approx 4\;\bar{\epsilon}\;\eta\,.
\label{eq:dlneps2}
\end{equation}
Since $\bar{\epsilon}$ is assumed to be nearly constant and large,
Eq.~(\ref{eq:dlneps2}) implies $|\eta|\ll 1$; that is, the potential
must be nearly exponential.

From the background solution given in Eqs.~(\ref{eq:exact}), it is
easily seen that
\begin{equation}
\tau\;H\;a \approx \left(\frac{1}{\bar{\epsilon}}\right)\cdot\{1+{\cal 
O}(\epsilon, \eta)\}\,.
\label{tau}
\end{equation}
Therefore, substituting Eqs.~(\ref{neweps}), (\ref{eq:dlneps2})
and~(\ref{tau}) into Eq.~(\ref{m2}), we find
\begin{equation}
\beta \approx 2(\epsilon+\eta)\,.
\label{m3}
\end{equation}

We may now proceed to calculate the spectral index of density perturbations.
As seen from Eq.~(\ref{eq:Phi}), the long-wavelength limit of the Newtonian 
potential is given by
\begin{equation}
k^{3/2}\Phi_k \sim k^{-n + 1/2}\approx k^{-\beta}\,,
\label{eq:lw}
\end{equation}
where we have used the fact that $n = \sqrt{\beta+1/4}$ and $\beta \ll 1$
in this case. 

Equation~(\ref{eq:lw}) describes the spectrum of $\Phi$ for modes that
went unstable during the contracting phase.  In Ref.~\cite{pert}, it
was argued that the pre-big bang spectrum of $\Phi$ gets imprinted on
the long-wavelength part of $\zeta$ as the universe undergoes reversal
from contraction to expansion.  (See also Refs.~\cite{khoury2},~\cite{tts}
and~\cite{durrer}.)  Then, the post-big bang spectrum of energy density
perturbations is given by $\delta_k\sim k^{-\beta}$, corresponding to
a spectral index
\begin{equation}
n_s - 1 \equiv \frac{d\ln|\delta_k|^2}{d\ln k} = -2\beta\,,
\label{eq:ns}
\end{equation}
where $n_s = 1$ corresponds to an exactly scale-invariant
(Harrison-Zel'dovich) spectrum.

Substituting Eq.~(\ref{m3}) into Eq.~(\ref{eq:ns}), we find that the
spectral index of density perturbations in the cyclic and ekpyrotic
scenarios is given by
\begin{eqnarray}
n_s - 1 = -4(\epsilon + \eta) =
-4\left\{\left(\frac{V}{V_{,\phi}}\right)^2 +  
1 - \frac{VV_{,\phi\phi}}{V_{,\phi}^2}\right\}\,.
\label{eq:nsek}
\end{eqnarray}
In the case of pure exponential potentials, $\eta$ vanishes
identically, and therefore the spectrum is red (since $\epsilon >0 $).
For potentials of larger curvature than an exponential, such as
$-e^{-c\phi^2}$, one has $\eta > 0$ and the spectrum is also red.
However, for potentials of smaller curvature than an exponential, such
as $e^{-c\sqrt{\phi}}$, one has $\eta < 0$, and the spectrum will be
blue if $\epsilon +\eta$ is also less than zero.  For instance, the
string-inspired potential of Ref.~\cite{orig} led to a blue spectrum.
We now see that both red and blue spectra can be achieved, as
anticipated by Linde {\it et al.}~\cite{linde}.
 
It is instructive to compare Eq.~(\ref{eq:nsek}) with its counterpart
in slow-roll inflation~\cite{wang,stewart}
\begin{equation}
n_s - 1 = -6\bar{\epsilon} + 2\bar{\eta}\,,
\label{eq:nsinf}
\end{equation}
where $\bar{\epsilon}= V_{,\phi}^2/2V^2$ and $\bar{\eta}=
V_{,\phi\phi}/V$ are the usual slow-roll parameters of inflation.  It
is easily seen that pure exponential potentials also yield a red
spectrum in this case.  Once again, it is possible to find potentials
for which the spectrum can be either red or blue.

In summary,  our work shows that the inflationary and recently
introduced ekpyrotic mechanisms are the complete set of 
 approaches for generating a nearly adiabatic,
scale-invariant spectrum of fluctuations from a single scale field
without extreme fine-tuning.
For the ekpyrotic case, we have shown how
the spectral index is related to fast-roll parameters that
characterize the slope and curvature of the scalar field potential
scale.  
In Ref.~\cite{khoury2}, we show that this constraint requires
essentially the same amount of fine 
tuning as the slow-roll conditions for inflation.
In Ref.~\cite{future}, we consider the spectrum in a mixed case where
the scalar field rolls from an expanding inflationary regime to a
contracting ekpyrotic regime.

We thank J. Erickson for helpful discussions.
This work was supported in part by NSERC of Canada (JK), and by US
Department of Energy grant DE-FG02-91ER40671 (SG and PJS).

\end{multicols}
\end{document}